\documentclass[12pt,aps,preprint,superscriptaddress]{revtex4}

\usepackage{amsmath,graphicx}
\usepackage{xcolor}
\usepackage{braket}
\usepackage{bm}
\usepackage{xcolor}	   
\newcommand{\parallelslant}{\mathbin{\!/\mkern-5mu/\!}}

\begin{document}
\title{Magnetoresistance from time-reversal symmetry breaking in topological materials}
	
\author{Jorrit C. de Boer}
\affiliation{MESA$^+$ Institute for Nanotechnology, University of Twente, The Netherlands}
\author{Denise P. Leusink}
\affiliation{MESA$^+$ Institute for Nanotechnology, University of Twente, The Netherlands}
\author{Alexander Brinkman}
\affiliation{MESA$^+$ Institute for Nanotechnology, University of Twente, The Netherlands}

\today
\begin{abstract}
Magnetotransport measurements are a popular way of characterizing the electronic structure of topological materials and often the resulting datasets cannot be described by the well-known Drude model due to large, non-parabolic contributions. In this work, we focus on the effects of magnetic fields on topological materials through a Zeeman term included in the model Hamiltonian. To this end, we re-evaluate the simplifications made in the derivations of the Drude model and pinpoint the scattering time and Fermi velocity as Zeeman-term dependent factors in the conductivity tensor. The driving mechanisms here are the aligment of spins along the magnetic field direction, which allows for backscattering, and a significant change to the Fermi velocity by the opening of a hybridization gap. After considering 2D and 3D Dirac states, as well as 2D Rashba surface states and the quasi-2D bulk states of 3D topological insulators, we find that the 2D Dirac states on the surfaces of 3D topological insulators produce magnetoresistance, that is significant enough to be noticable in experiments. As this magnetoresistance effect is strongly dependent on the spin-orbit energy, it can be used as a telltale sign of a Fermi energy located close to the Dirac point.
\end{abstract}

\maketitle

\section{Introduction}	
\label{sec:helical_MR}
It is well known that magnetoresistance effects can often be described in terms of Shubnikov-de Haas quantum oscillations and Drude multiband magnetoresistance and that this can be used to gather detailed information about the electronic structure of a material. However, these effects do not always fully describe the physics at hand and magnetoresistance may arise through other mechanisms. For instance, there are many reports of large magnetoresistance in Bi-based and Heusler topological insulators (TIs) \cite{he_high-field_2012,qu_quantum_2010,tang_two-dimensional_2011,wang_room_2012,wang_large_2013,wang_granularity_2014,breunig_gigantic_2017}, which are difficult to explain using the simplified Drude model and require one to look into different sources of large magnetoresistance. In 1969, Abrikosov derived the occurence of large, linear magnetoresistance for cases where only the lowest Landau level is filled \cite{abrikosov_galvanomagnetic_1969,abrikosov_quantum_2000}. To observe this effect, the system needs to be in the quantum limit: $E_F, k_B T \ll \delta E_{LL}$, where $\delta E_{LL}$ is the energy difference between two successive Landau levels and $E_F$ and $k_B T$ represent the Fermi and thermal energies, respectively. This can usually only be fulfilled at extremely low carrier densities and high electron mobilities, as is the case for Bi \cite{kapitza_p._study_1928} and n-type doped InSb \cite{hu_classical_2008}. Because of the lower mobilities in Bi-based topological insulators, quantum linear magnetoresistance seems unlikely to occur in these systems and the large magnetoresistance has to originate from another mechanism. On the other hand, in very disordered systems, classical magnetoresistance has been predicted\cite{parish_classical_2005,parish_non-saturating_2003}. In this work, we will focus on the intermediate regime and discuss the magnetoresistance that is already embedded inside the Zeeman term in model Hamiltonians that describe Bi-based topological materials with relatively low mobilities. 

\section{Helical magnetoresistance}
\label{sec:HMR}
The approximations within the Drude model do not only make life easier, they also neglect effects that may be very useful for characterizing the electronic structure. For example, the charge carrier mobility $\mu = e \tau m^{-1}$ (with $\tau$ the scattering time and $m$ the effective mass) does not have to be constant with field and $\rho_{\textrm{xx}} = m / (n e^2 \tau)$, where $n$ is the charge carrier density, can aquire a magnetic field dependence through the scattering rate $\Gamma(B) = \tau^{-1}(B)$. In the following, we will investigate how the magnetic field dependence of the scattering time influences the magnetoresistance of TIs and related systems with strong spin-orbit coupling.

\subsection{Surface Dirac cones}
The magnetic field couples to an electronic system through two main mechanisms: the Zeeman effect and the 'orbital' or 'Doppler' effect $\bm{p}^{\prime}=\bm{p}+e\bm{A}$, where $\bm{p}$ is the electron momentum and $\bm{A}$ the vector potential. Here, we focus on topological insulators with low mobilities such that $\omega_c \tau \ll 1$ ($\omega_c$ represents the cyclotron frequency and $\tau$ the scattering time) and the influence of the orbital effect is small, as is the case for typical TI thin films. Ignoring the orbital effect of a magnetic field, topological surface states of Bi-based 3D topological insulators can be modeled using the 2D Hamiltonian by Liu \textit{et al.} \cite{liu_model_2010}:
\begin{equation}
H_{TSS} = \hbar v_F (\bm{\sigma} \times \bm{k}) + \frac{g \mu_B}{2} \, \sigma_z B_z, 
\label{eq:TSS}
\end{equation}
where $\mu_B$ is the Bohr magneton, $g$ is the effective magnetic moment and $\bm{\sigma}$ is the vector containing the 3 Pauli matrices to represent the spin degree of freedom. Note that the spin-orbit interaction part of the Hamiltonian is essentially the Rashba Hamiltonian $H_{RSOC}=\frac{\alpha}{\hbar} (\bm{\sigma} \times \mathbf{p}) \cdot \mathbf{e_z}$, with $\alpha$ indicating the spin-orbit coupling strength. Due to this spin-orbit interaction, the degenerate energy bands have opposite helicities, which are denoted by the $\pm$ indices in the following. The Zeeman effect, arising trom a magnetic field in the $z$-direction, is captured by a Hamiltonian of the simple form  $H_Z = (g \mu_B / 2) \, \bm{\sigma} \cdot \bm{B}$, which describes the alignment of the spins in the magnetic field direction.

\begin{figure}
\includegraphics[clip=true,width=0.7\textwidth]{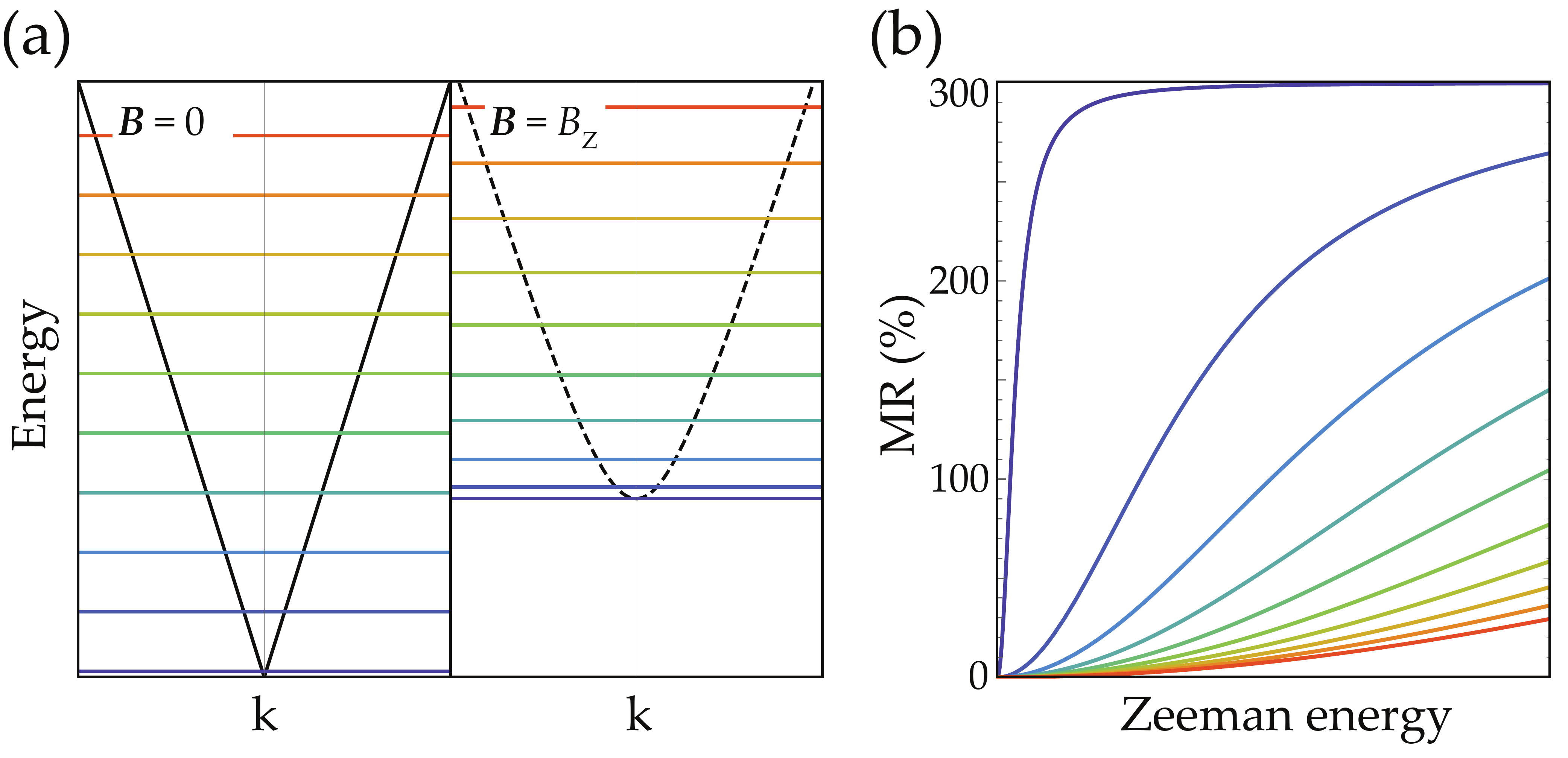}
\centering
\caption{\textbf{Helical magnetoresistance.} \newline (a) 3D TI surface Dirac cones (black lines) in the absence (left panel) and presence (right panel) of a magnetic field perpendicular to the 2-dimensional electron gas (2DEG). Horizontal lines of the same color indicate how the Fermi energy changes with magnetic field, keeping the carrier density $n_{2D} = k_F^2/(2\pi)$ constant. (b) Helical magnetoresistance as a function of Zeeman energy for different initial Fermi energies, i.e. different spin-orbit energies, where purple (red) corresponds to small (large) $E_{\textrm{SO}}$.}
\label{fig:HelicalMR}
\end{figure}

Writing $|\bm{k}| = k$, the dispersion relation of the conduction band side of the system is given by 
\begin{equation}
E_C = \sqrt{\hbar^2 v_F^2 k^2 + (g \mu_B B_z / 2)^2}
\end{equation}
with the corresponding spinors
 \begin{equation}
 	\psi_{C,\pm} = \frac{1}{\sqrt{2 E_C}}\begin{pmatrix} i e^{-i\theta} \sqrt{E_C + g \mu_B B_z/2} \\
 							\pm \sqrt{E_C - g \mu_B B_z/2} 
 							\end{pmatrix}
\label{eq:2Dcone_spinors}
 \end{equation}
for the top and bottom surfaces of the TI. 
Within a simple Boltzmann picture, the scattering rate $\Gamma_B = \tau_{B}^{-1}$ is proportional to the number of available states to scatter to. Assuming dominant elastic scattering, the scattering rate is given by an integral over the Fermi surface: $\tau_{B}^{-1} \propto \int {S (1- \cos \theta ) d \theta}$, where the scattering factor $S$ is determined using Fermi's golden rule, $S_\pm = |\Braket{\psi_\pm ^{\prime} | \psi_\pm}|^2$, for scattering from $\left| \psi \right>$ at zero angle to $\left| \psi ^{\prime} \right>$ at angle $\theta$. For scattering within a single Dirac cone we find 
\begin{equation}
S_{+} = S_{-} = \frac{\frac{1}{2} (1+ \cos \theta ) \hbar^2 v_F^2 k^2 + (g \mu_B B_z/2)^2 }{ \hbar^2 v_F^2 k^2 + (g \mu_B B_z/2)^2 }.
\end{equation} 
This expression reduces to $\frac{1}{2}(1+\cos \theta)$ for $B \rightarrow 0$, which describes the well known suppressed backscattering in TIs \cite{li_4pi-periodic_2018}, induced by the helical spin ordering. Through $\langle \psi_{C,\pm} | \sigma_z | \psi_{C,\pm} \rangle$, we find the out-of-plane component of the spin to be $S_z = \frac{\hbar}{2} (E_z / E_{C})$, where we used $E_z=g \mu_B B_z/2$. For nonzero magnetic field, the helical order is broken as all spins are tilted along the magnetic field direction, creating a finite overlap between states in every momentum-space direction, which allows backscattering. A compact expression for the dependence of the scattering rate on the magnetic field (and therefore for the magnetoresistance) is found by multiplying the scattering factor $S_\pm$ with the Boltzmann factor $(1-\cos\theta)$ and integrating the result over all angles $\theta$. We find for the magnetoresistance:
\begin{equation}
MR_{Helical} = 100\% \times \frac{R(B)-R(0)}{R(0)} \propto 100\% \times \frac{3 x^2}{1+x^2},
\label{eq:MR}
\end{equation}
where $x$ is given by $x(B)= E_\textrm{Z}(B) / E_{\textrm{SO}}$ and can be seen as a competition between the Zeeman energy $E_\textrm{Z}=g \mu_B B /2$ and the spin-orbit energy at the Fermi level $E_{\textrm{SO}}= \hbar v_F k_F$. The difference between the zero field limit and the large field limit results in a magnetoresistance of 300\%. This factor 4 difference in transport scattering time between the cases of spin-momentum locked spins and fully aligned spins, was first pointed out by Wu \textit{et al.} \cite{wu_sudden_2013}.

Fig \ref{fig:HelicalMR} illustrates the effect of the Zeeman energy on the band structure and magnetoresistance. Fig \ref{fig:HelicalMR}(a) shows the evolution of the Fermi level with increasing Zeeman energy. Because we assume the carrier density $n_{2D} = k_F^2/(2\pi)$ to be constant, the spin-orbit energy $E_{\textrm{SO}}= \hbar v_F k_F$ remains unaffected by the magnetic field. Note that the opening of a gap with magnetic field is not an additional effect, but a visualization of the hybridization term that causes the enhanced scattering probability. In Fig \ref{fig:HelicalMR}(b) the Zeeman energy and thereby the ratio $x(B)= E_\textrm{Z}(B) / E_{\textrm{SO}}$ is varied for different spin-orbit energies. From this figure, we see that especially for Fermi levels close to the Dirac point, the magnetoresistance through broken spin helicity quickly reaches its saturation value of 300\%. 

For a realistic g-factor of $25$ \cite{liu_model_2010} and a magnetic field of 10~T, we can substitute $E_{\textrm{SO}}^2 = E_{\textrm{Z}}^2 - E_F^2$ (with $E_F$ the Fermi energy) into Eqn. (\ref{eq:MR}) and find that to reach a 100\% helical MR, the Fermi level needs to be within $\sim$ 10~meV with respect to the Dirac point. While this effect is strong enough to survive thermal broadening at liquid Helium temperatures, inhomogeneities in the electronic structure of the Bi-based TIs may smear out the effect over a larger energy range.

\subsection{Rashba-type surface states}
In the previous section we have seen that in non-degenerate, surface Dirac cones, described by a Hamiltonian that is dominated by Rashba-type spin-orbit coupling, large MR up to $\sim$ 300\% can arise. In this section, we study the response of Rashba-type surface states to a magnetic field. Apart from a large parabolic contribution to the band structure, the system is described by spin-orbit coupling that causes spin-momentum locking in a similar fashion as in the 2D TI surface Dirac cone. So to model Rashba surface states, we use a similar model Hamiltonian as in the previous section, but here the Rashba and magnetic field parts act as corrections to a dominant parabolic term:  
\begin{equation}
H_{RSS} = \frac{\hbar^2 k^2}{2 m^*} + \hbar v_F (\bm{\sigma} \times \bm{k}) + \frac{g \mu_B}{2} \, \sigma_z B_z.
\label{eq:RSS}
\end{equation}
Here, the resulting dispersion relations
\begin{equation}
E_{C,\pm} = \frac{\hbar^2 k^2}{2 m^*} \pm \sqrt{\hbar^2 v_F^2 k^2 + (g \mu_B B_z/2)^2}
\end{equation}
both correspond to conduction band states on the same surface, but with opposite helicities. The spinors of these two conduction bands are: 
 \begin{equation}
 	\psi_{C,\pm} = \frac{1}{\sqrt{2 E_{C,\pm} - \frac{\hbar^2 k^2}{2 m^*}}} \begin{pmatrix} i e^{-i\theta} \sqrt{E_{C,\pm} + g \mu_B B_z/2} \\
 							\pm \sqrt{E_{C,\pm} - g \mu_B B_z/2} 
 							\end{pmatrix},
\label{eq:Rashba_spinors}
 \end{equation}
which is very similar to the Dirac cone spinors of equation \ref{eq:2Dcone_spinors}. The apparently small, but very important difference, is the use of different energy disperions $E_{C,\pm}$ for the two spinors. In this case, the out-of-plane component of the spin $S_z = \frac{\hbar}{2} \, E_z / [E_{C,\pm} - \hbar^2 k^2 /(2 m)]$, which tells us that in high magnetic fields, the spins of the two helicities align in opposite directions along the $k_z$-axis. 

For the Rashba 2DEG, the total amount of available states to scatter to, doubles with respect to the single Dirac cone and the scattering factor becomes $S_\pm = | \langle \psi_+^{\prime} | \psi_\pm \rangle |^2 + | \langle \psi_- ^{\prime} | \psi_\pm \rangle |^2$. Because of interband scattering, we find for zero field: $S_\pm = \frac{1}{2}(1+\cos{\theta}) + \frac{1}{2}(1-\cos{\theta}) = 1$. In the high field limit, intraband scattering becomes possible at all angles $\theta$ and $| \langle \psi_+ ^{\prime} | \psi_\pm \rangle |^2 \rightarrow 1$. However, because of the opposite magnetic field response of the two helicities in the Rashba system, interband scattering becomes strongly suppressed in the high field limit ($| \langle \psi_- ^{\prime} | \psi_\pm \rangle |^2 \rightarrow 0$), so that $S_\pm \rightarrow 1$, which is the exact same result as for zero field. We conclude that in contrast to the single Dirac-type surface state, the set of two Rashba-type surface states results in zero net helical magnetoresistance.

\section{Magnetoresistance through a change in Fermi velocity}
Upon following textbook derivations of the Drude resistance from the Boltzmann transport equation, but now for a single, spin non-degenerate band and without assuming a parabolic dispersion relation, one arrives at
\begin{equation}
		\rho_{\textrm{xx}} = \frac{4\pi}{k_F^2}\frac{\hbar k_F/v_F}{e^2 \tau}.
		\label{eq:rhoxx}
\end{equation} 
While the substitutions $4\pi/k_F^2 \rightarrow n_{2D}$ and $\hbar k_F/v_F \rightarrow 1/m^{*}$ recover the Drude model for parabolic bands, Eqn. \ref{eq:rhoxx} indicates $k_F^{-1}$ and $v_F^{-1}$ dependencies of the magnetoresistance. In this section we consider the effect of the opening of a Zeeman gap (as in Fig \ref{fig:HelicalMR}(a)) on the Fermi velocity and the resulting magnetoresistance, while we assume the carrier density - and therefore $k_F$ - to be constant.

\subsection{Surface Dirac cones}
Considering the model Hamiltonian for 2D Dirac surface states, Eqn. (\ref{eq:TSS}), the Fermi velocity changes with magnetic field as
\begin{equation}
v_F(B) = \frac{1}{\hbar} \frac{ \partial}{\partial k} \sqrt{\hbar^2 v_F^2 k^2 + (g \mu_B B_z/2)^2} = \frac{v_F E_\textrm{SO}}{\sqrt{E_\textrm{SO}^2 + E_\textrm{Z}^2}},
\end{equation} 
so that $\rho_{\textrm{xx}} \propto \sqrt{E_\textrm{SO}^2 + E_\textrm{Z}^2}/E_\textrm{SO}$. Then we find an additional magnetoresistance originating from a change in Fermi velocity as the bands aquire a Zeeman-shift: 
\begin{equation}
MR_{v_F} = 100\% \times (\sqrt{1+x^2}-1),
\label{eq:velocity_MR}
\end{equation}
where we still use $x(B)= E_\textrm{Z}(B) / E_{\textrm{SO}}$. We see that the decrease of Fermi velocity with increasing magnetic field causes a non-saturating magnetoresistance, which becomes linear in $B$ in the high-field limit $ E_\textrm{Z}(B) \gg E_{\textrm{SO}}$. Including the magnetic field dependencies of both the scattering time and Fermi velocity, we obtain an expression for the Zeeman-induced magnetoresistance in 2D Dirac surface states:
\begin{equation}
MR_{v_F,Helical} =  100\% \times \left[ \left( 1 + \frac{3x^2}{1+x^2} \right) \sqrt{1+x^2} - 1\right].
\label{eq:total_Dirac_MR}
\end{equation}
Through this model, as $x \propto B \rightarrow \infty$, enormous magnetoresistance values can be reached for low carrier densities (i.e. Fermi energies close to the Dirac point) as in this regime the resulting magnetoresistance becomes linear. Comparing our findings with experimental results, we note that linear magnetoresistance is very common in measurements on topological surface states\cite{he_high-field_2012,qu_quantum_2010,tang_two-dimensional_2011,wang_room_2012,wang_large_2013,wang_granularity_2014,breunig_gigantic_2017}. However, distinguishing the described effect from classical magnetoresistance arising from strong inhomogeneity\cite{parish_classical_2005,parish_non-saturating_2003} may be difficult.

\subsection{Rashba-type surface states}
For the Rashba-type surface states described by Eqn. (\ref{eq:RSS}), the Fermi velocity dependence on the magnetic field should be significantly less dramatic as in this case the dispersion relation is dominated by the parabolic term. Following the same procedure as above (and assuming a fixed $k_F$ for simplicity), we find the magnetoresistance as a consequence of the Fermi velocity change in a single, spin non-degenerate band to be:
\begin{equation}
MR_{v_F} = 100\% \times \frac{\sqrt{1+x^2}-1}{1 \pm \frac{2 E_\textrm{p}}{E_\textrm{SO}}\sqrt{1+x^2}},
\label{eq:velocity_MR2}
\end{equation}
where $E_\textrm{p} = \hbar^2 k^2 /(2 m^*)$ is the parabolic contribution to the dispersion. It is instructive to consider this result in a few limits. In the high field limit $E_\textrm{Z} \gg E_{\textrm{SO}}$ ($x\gg1$), we can further explore the limits $ E_\textrm{p} \gg E_{\textrm{SO}} $ and $ E_\textrm{p} \ll E_{\textrm{SO}} $:
\begin{align}
MR_{v_F} 	&=  100\% \times \frac{x-1}{1 \pm \frac{2 E_\textrm{p}}{E_\textrm{SO}} x} \\
			&\rightarrow  
	\begin{cases}
		\textrm{Saturation at }\pm\frac{E_{\textrm{SO}}}{2 E_\textrm{p}} &\textrm{for } E_\textrm{p} \gg E_{\textrm{SO}} \\ 
		\textrm{Linear MR} &\textrm{for } E_\textrm{p} \ll E_{\textrm{SO}} 
	\end{cases}.
\end{align}
Note that in the limit $E_\textrm{p} \gg E_{\textrm{SO}}$, the MR contributions from the two individual $E_{C,\pm}$ bands are opposite. Without correctly summing the conductivity, this already hints at cancelling contributions that result in zero net effect.  
\begin{figure*}
\includegraphics[clip=true,width=1\textwidth]{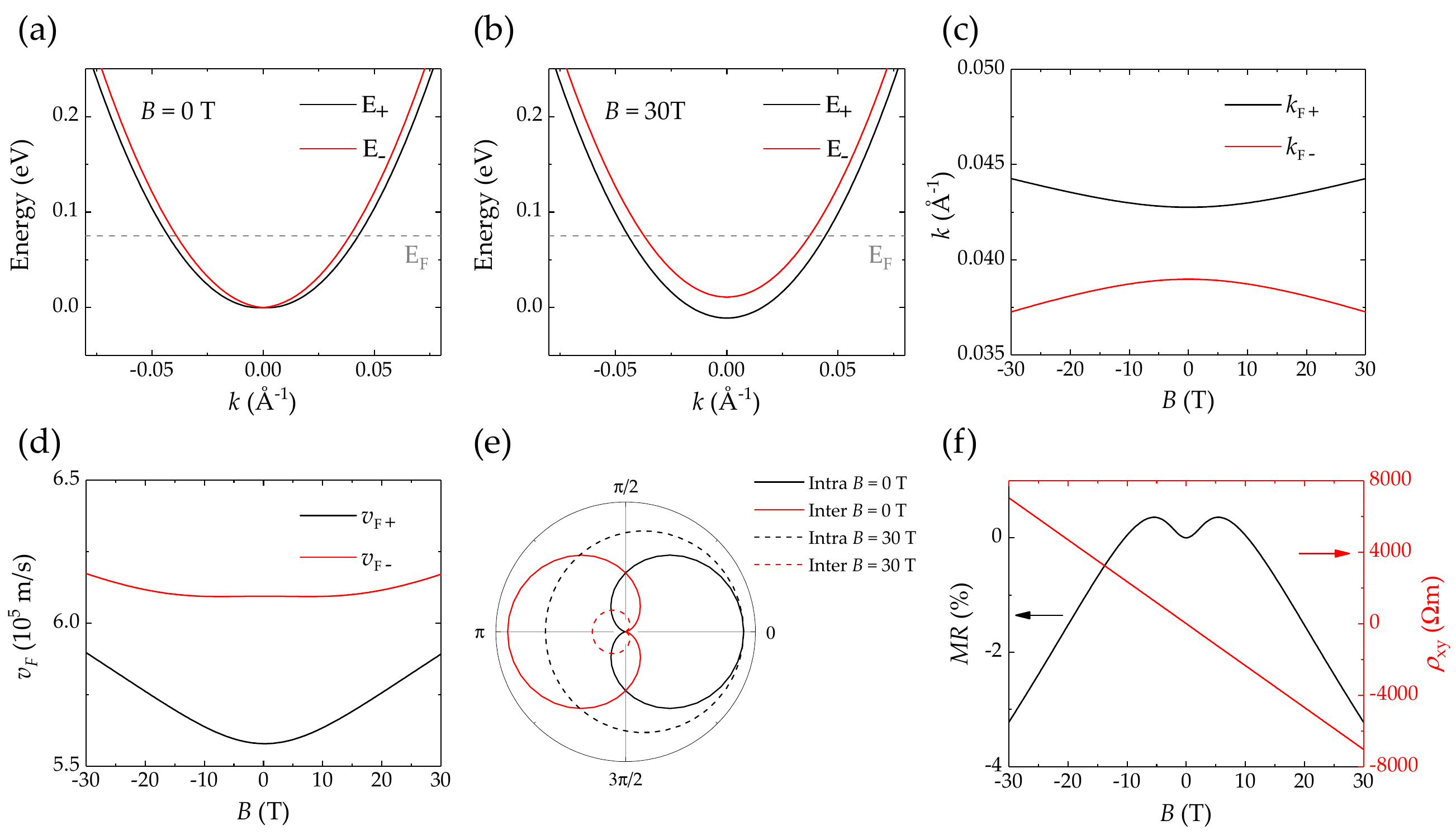}
\centering
\caption{\textbf{Rashba 2DEG model in a perpendicular magnetic field.} \newline (a) The Rashba 2DEG dispersion without magnetic field. (b) Dispersion in a $30$T magnetic field. (c) Evolution of the wave number $k$ with applied magnetic field. (d) Evolution of the Fermi velocity with magnetic field. (e) Intra- and interband wavefunction overlap of $\psi(\theta=0)$ with other states on the Fermi surface, for $B = 0$T and $B = 30$T. The radius indicates the wavefunction overlap and the forward directed state marks an overlap of $1$. (f) Magnetoresistance (black line) and Hall resistivity (red line) arising from the response of the Fermi wave vectors, Fermi velocities and scattering times to a perpendicular, external magnetic field.}
\label{fig:BTE_num_data}
\end{figure*}

Because of the complexity that arises when the conductivity contributions from both bands are summed and matrix-inverted to resistivity, we resort to numerical methods from here on. In the numerical model we use the following parameters: $E_F = 75$~meV, $\hbar v_F = 0.17$~eV~$\textrm{\AA}$, $\hbar^2/(2 m) = 45$~eV~$\textrm{\AA}^2$ and $g = 12$, which do not represent a specific material, but are comparable to the Rashba-like states in Bi-based TIs \cite{liu_model_2010}. In figure \ref{fig:BTE_num_data}, we present several results from the model. Figs~\ref{fig:BTE_num_data}(a) and (b) illustrate the band structures without and with magnetic field respectively. Most apparent from these figures is the splitting of the two bands due to the magnetic field, which changes the carrier densities for the different helicities. The latter is also clear from Fig \ref{fig:BTE_num_data}(c), where we see that the total carrier density is conserved. From Fig \ref{fig:BTE_num_data}(d), we see that the simplification from earlier, that $k_F \approx$ constant, caused us to miss a change in total Fermi velocity with magnetic field for this Rashba system. This small increase of $v_F$, unaffected by the constant scattering time (see panel \ref{fig:BTE_num_data}(e)), results in a small, negative magnetoresistance $MR \approx -3\%$ as shown in Fig \ref{fig:BTE_num_data}(f). From this, we can conclude that in 2D Rashba surface states, no noteworthy magnetoresistance arises through the magnetic field dependent scattering time, Fermi velocity or even a combination of the two.

\section{Magnetotransport through the bulk of a 3D topological insulator}
\begin{figure}
\includegraphics[clip=true,width=0.5\textwidth]{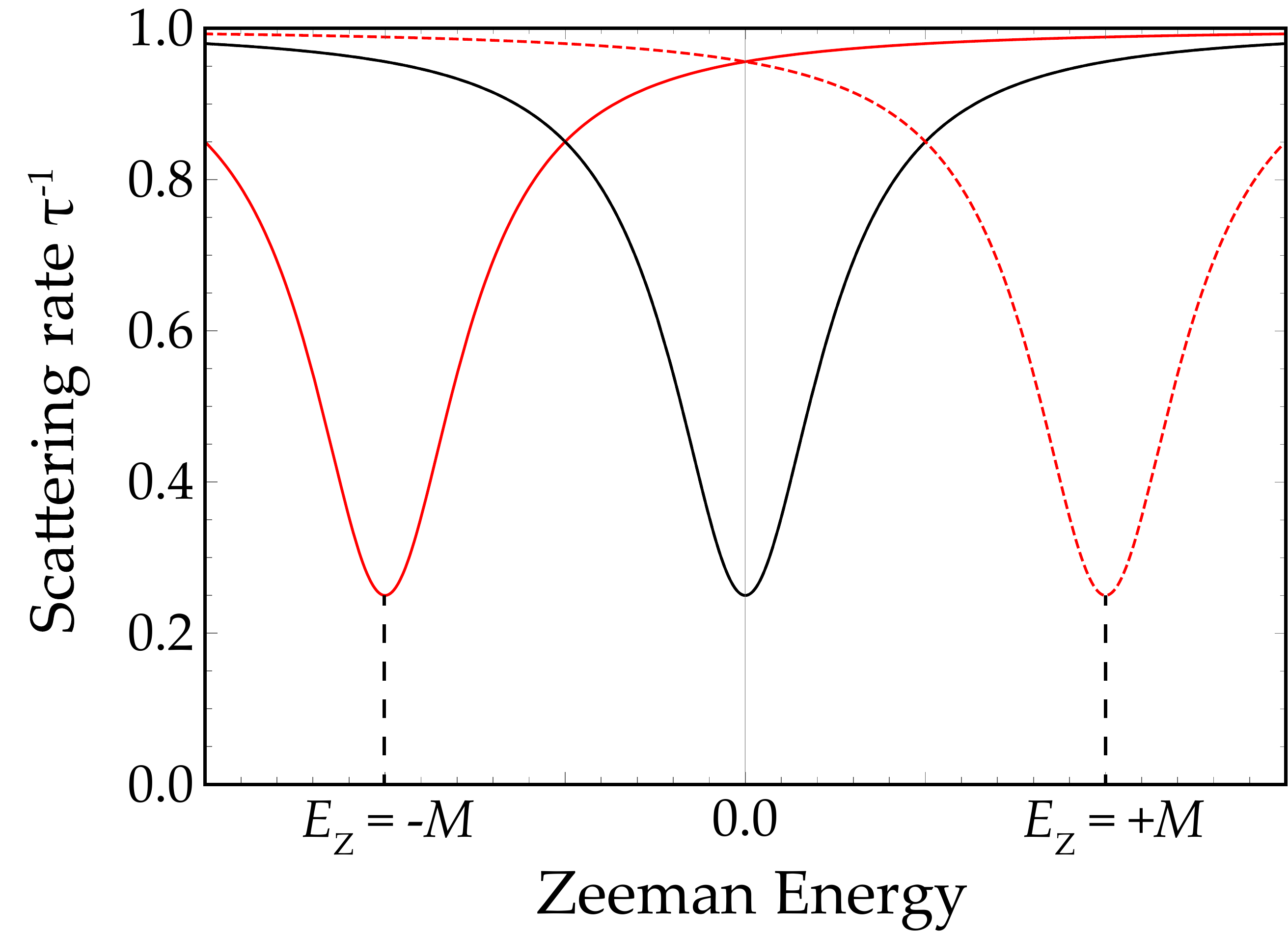}
\centering
\caption{\textbf{Mass-term dependence of the scattering rate.} \newline The black line represents the magnetic field dependence of the scattering rate for $M \rightarrow 0$ and shows the recovery of the factor $4$ from the surface Dirac cone. The normalized scattering rate corresponds to the normalized wavefunction overlap. The solid(dashed) red line indicates the scattering rate of the $+$($-$) helicity for nonzero mass term $M$. The mass term acts as an offset to the Zeeman term, but in opposite directions for the different helicities.}
\label{fig:MR_effect_of_bulk_M}
\end{figure}

In the Bi-based TI family, there are only few examples of alloys that are true bulk insulators and the majority exhibits a bulk shunt \cite{hasan_colloquium:_2010,qi_topological_2011}. To describe the bulk states, we once more utilize the work horse bulk Hamiltonian from Liu \textit{et al.} \cite{liu_model_2010}. Up to \textit{O}$(k^2)$, rotated around the $y$-axis in orbital space ($\sigma_x \leftrightarrow \sigma_z$) and around the $z$-axis in spin space ($\theta \rightarrow \theta + \pi$) it reads:
\begin{equation}
H_{\textrm{Liu}} = E^0_{\bm{k}} \, \sigma_0 s_0 + M_{\bm{k}} \, \sigma_x s_0 + \hbar v_{\parallelslant} \, \sigma_z(s_y k_x - s_x k_y) + \hbar v_z k_z \,\sigma_y s_0 + (g \mu_B / 2) B_z\, \sigma_0 s_z , 
\label{eq:HLiuBIG}
\end{equation}
where $E^0_{\bm{k}}$ and $M_{\bm{k}}$ are polynomials in $k_{\parallelslant}$ and $k_z$. In principle, Eqn. (\ref{eq:HLiuBIG}) describes two Rashba systems of opposite sign, coupled by $M_{\bm{k}}$ and $\hbar v_z k_z$. As in these materials the dispersion in the $z$-direction is allmost negligible, $v_z$ is much smaller than $v_{\parallelslant}$  \cite{liu_model_2010}. $M_{\bm{k}}$ however, is not necessarily small and we continue with the $4\times4$ Hamiltonian, where we neglect $\hbar v_z k_z$ and the parabolic $E^0_{\bm{k}}$ term for simplicity. Taking $E_{\textrm{SO}}= \hbar v_F k_F$ and $E_\textrm{Z}=g \mu_B B/2$, we find for the conduction band two dispersions,
\begin{equation}
E_{C,\pm} = \sqrt{E_{\textrm{SO}}^2 + (M_{\bm{k}} \pm E_{\textrm{Z}})^2},
\end{equation}
with the spinors
\begin{equation}
 	\psi_{C,\pm} = \frac{1}{\sqrt{A_0}} \begin{pmatrix} E_{\textrm{SO}} \\
 							(E_{C,\pm} \mp M_{\bm{k}} - E_{\textrm{Z}} ) \, i e^{i\theta} \\
 							\pm E_{\textrm{SO}} \\
 							\mp (E_{C,\pm} \mp M_{\bm{k}} - E_{\textrm{Z}} ) \, i e^{i\theta}
 							\end{pmatrix},
\label{eq:Bulk_spinors}
 \end{equation}
where $A_0 = 4 E_{C,\pm} (E_{C,\pm} \mp M_{\bm{k}} - E_{\textrm{Z}})$ is the normalization factor.
These two spinors are orthogonal for every angle in momentum-space, so that interband scattering is forbidden in the bulk conduction band. Using $S_\pm =  | \langle \psi_{\pm} ^{\prime} | \psi_\pm \rangle |^2$, we find for the helical magnetoresistance
\begin{equation}
MR_{H,\pm} \propto \frac{100\% \times 3 \, E_\textrm{SO}^2 E_Z (E_Z \pm 2 M)}{(E_\textrm{SO}^2 + 4 M^2)(E_\textrm{SO}^2+(E_Z \pm M)^2)},
\label{eq:bulk_helical_MR}
\end{equation}
where $M = M_0$ is the momentum-independent part of $M_{\bm{k}} = M_0 + M_1 k_{\parallelslant}^2 + M_2 k_z^2$ and represents the gap size. The $\pm$ sign indicates that the mass term acts as an offset to the magnetic field term. In Fig \ref{fig:MR_effect_of_bulk_M}, it is shown that the offset due to finite $M$ significantly reduces the effect of magnetic fields that are small with respect to $M$ (as is the case for Bi-based TIs). Moreover, the opposite response of the two helicities to the magnetic field causes the MR from the separate bands to cancel. So while interband scattering is forbidden in the TI bulk (which suppressed helical MR in Rashba surface states), it is the gap $M$ that makes the helical MR effect small. 

Similar to the $2$D Rashba system, the Zeeman-shift works in opposite ways for the Fermi velocities of the two helicities,
\begin{equation}
v_F(B) = \frac{1}{\hbar} \frac{ \partial}{\partial k} \sqrt{\hbar^2 v_F^2 k^2 + (g \mu_B B_z/2)^2} \propto \frac{\alpha E_\textrm{SO}}{E_\textrm{SO}^2 + (M \pm E_Z)^2}.
\end{equation} 
As a consequence, also the correction to the Fermi velocity by the Zeeman-shift does not cause any magnetoresistance in the bulk of topological insulators, similar to the case for 2D Rashba states. 

In the limit $M \rightarrow 0$, Eqn. (\ref{eq:HLiuBIG}) describes an accidental DSM, with two linear, orthogonal Dirac cones. Because of the 3D character, we should also take the $k_z$-dependence into account and use
\begin{equation}
H_{\textrm{Liu}}^{DSM} = \hbar v_{\parallelslant} \, \sigma_z(s_y k_x - s_x k_y) + \hbar v_z k_z \,\sigma_y s_0 + (g \mu_B/2) B_z\, \sigma_0 s_z.  
\label{eq:HLiuDSM}
\end{equation}
In spherical coordinates and in terms of $E_{\textrm{SO}\parallelslant} = \hbar v_{\parallelslant} k_{\parallelslant}$, $E_{\textrm{SO}\perp} = \hbar v_z k_z$ and $E_{\textrm{Z}} = g \mu_B B_z/2$, the dispersion of the conduction band 
\begin{equation}
E_{C,\pm} = \sqrt{E_{\textrm{SO} \parallelslant }^2 \, \sin{\varphi}^2 + ( E_{\textrm{SO}\perp} \, \cos{\varphi} \pm E_{\textrm{Z}})^2} \end{equation}
and the spinor parts of the wavefunctions become
\begin{equation}
 	\psi_{C,\pm} = \frac{1}{\sqrt{A_{\pm}}} \begin{pmatrix} 
 	
 							e^{i\theta} \sin{\varphi} \, E_{\textrm{SO}\parallelslant} \\
 							i (E_{\textrm{Z}} \pm \cos{\varphi} \, E_{\textrm{SO}\perp} - E_{C,\pm} )  \\
 							e^{i\theta} \sin{\varphi} \, E_{\textrm{SO}\parallelslant} \\
 							\pm E_{\textrm{Z}} + \cos{\varphi} \, E_{\textrm{SO}\perp} \mp E_{C,\pm}
 	\end{pmatrix},
\label{eq:DSM_spinors}
 \end{equation}
with the normalization factor $A_{\pm} = 2 \sin{\varphi}^2 \,E_{\textrm{SO}\parallelslant}^2 + 2 (\pm E_{\textrm{Z}} + \cos{\varphi} \, E_{\textrm{SO}\perp} \mp E_{C,\pm})^2$. As was the case for the bulk 3D TI spinors of the last section, the two spinors for the conduction band side of $H_{\textrm{Liu}}^{DSM}$ are completely orthogonal. Note that for $\varphi = \pi/2$, we recover a two-fold degenerate version of the $2$D surface Dirac cone system used in the above, which indicates that large, helical magnetoresistance may be present in this system. However, the $3$D character of the DSM allows the magnetic field term to be just absorbed into $k_z^{\prime} = k_z \pm g \mu_B B / (\hbar v_F)$ and the Dirac system simply splits into two, ungapped Weyl cones. Not only does the absence of a gap discard the effect of the Zeeman-shift on the Fermi velocity, it also means that the branches are not hybridized and that even in high magnetic fields, direct backscattering is still not possible within this linearized model. Therefore, $3$D Dirac semimetals should be free of both helical and Zeeman-shift induced magnetoresistance.

\section{Conclusions}
In this work, we studied how magnetotransport in topological materials can originate directly from a generic TI Hamiltonian with a Zeeman term. We found that the experimentally observed large magnetoresistance in Bi-based 3D topological insulators \cite{he_high-field_2012,qu_quantum_2010,tang_two-dimensional_2011,wang_room_2012,wang_large_2013,wang_granularity_2014} can partially be explained by detailed effects incorporated in the model Hamiltonians. While we found no significant contributions to the magnetoresistance by topological bulk or surface Rashba states (apart from possibly causing multiband magnetoresistance), surface Dirac cones can cause large, non-saturating, linear magnetoresistance through both the scattering time via broken time reversal symmetry and a correction to the Fermi velocity by means of a Zeeman-shift. As these effects are the largest when the Zeeman energy is of the same order as the spin-orbit energy, a large, non-saturating magnetoresistance may be a telltale sign of a Fermi level very close to the Dirac point.

\begin{acknowledgments}
This work was financially supported by the European Research Council (ERC) through a Consolidator Grant.
\end{acknowledgments}

\bibliographystyle{plain}
\bibliography{references}

\end{document}